\documentclass[a4paper,11pt,oneside,reqno]{article}
\usepackage{a4,amsmath,amssymb,latexsym,graphicx}

\begin{document}

\vskip 4cm

\begin{center}
{\LARGE Critique of an Experiment Regarding Quantum States\\

\vskip 0.05cm

in a Gravitational Field and Possible\\

\vskip 0.05cm

Geometrical Explanations\\}

\vskip 1cm {\large D.
Olevik$^{\dagger,}$\footnote{davole-8@student.luth.se}, C.
T\"{u}rk$^{\dagger,}$\footnote{chrtur-8@student.luth.se}, H.
Wiklund$^{\dagger,}$\footnote{hanwik-7@student.luth.se}, J.
Hansson$^{\ddagger,}$\footnote{hansson@mt.luth.se} }
\end{center}

\vskip 0.5cm

\begin{center}
Division of Physics\\Lule\aa \ University of Technology\\
SE-97187 Lule\aa\\ Sweden

\vskip 0.5cm

$^{\dagger}$ {\it Student project group} \ \ $^{\ddagger}$ {\it
Supervisor of project}
\end{center}

\vskip 0.5cm

\begin{abstract}
We discuss an experiment conducted by Nesvizhevsky et al. As it is
the first experiment claimed to have observed gravitational
quantum states, it is imperative to investigate all alternative
explanations of the result. In a student project course in applied
quantum mechanics, we consider the possibility of
quantummechanical effects arising from the geometry of the
experimental setup, due to the "cavity" formed. We try to
reproduce the experimental result using geometrical arguments
only. Due to the influence of several unknown parameters our
result is still inconclusive.
\end{abstract}

\newpage
\tableofcontents
\newpage

\section{Introduction}
A wellknown property of quantum mechanics is the quantisation of
the energy levels of a particle trapped in a potential well. For
instance, the electromagnetic and the strong nuclear force create
different kinds of potential wells and are responsible for many
observed phenomena in nature, such as the structure of atoms and
nuclei. This suggests that a splitting of the energy levels should
also be observed for particles in the Earth's gravitational field.
But, since the gravitational field is much weaker, the effect
should be subtle and hard to detect.

In a letter to Nature \cite{Nev1}, Nesvizhevsky et al. claim to
have observed such quantum effects of gravity acting on ultracold
neutrons (UCNs). They conducted an experiment in which the UCNs
were allowed to flow through a cavity with a reflecting surface
below and an absorber above. By measuring the number of neutrons
exiting the experimental setup, they claim to have observed
discrete energy levels. However, in their argument they seemingly
disregard the modification due to the absorber, stating that it is
``sufficiently perfect''. Further they argue that the
discretisation is related to the sudden increase of neutrons
coming through at distinct widths between the reflecting surface
and absorber. However, since the UCNs are restricted by both the
reflecting surface and the absorber, also the geometrical effects
should be considered. The results might even be explained by
geometrical arguments only.

The aim with this report is to show that by only using geometrical
arguments, the effects observed by the scientist at Grenoble can
be explained.

\section{Background}
Here we give a brief review of the experiment reported in
\cite{Nev1}. A similar experiment was first suggested by V.I.
Luschikov and A.I. Frank in 1978 \cite{Qeffect78}. We also discuss
inconsistencies of their theoretical analysis and their results,
giving us some ideas of how to approach an alternative
explanation.

\subsection{The experiment}
\begin{figure}[!htbp]
  \centering
  \includegraphics[width=13cm]{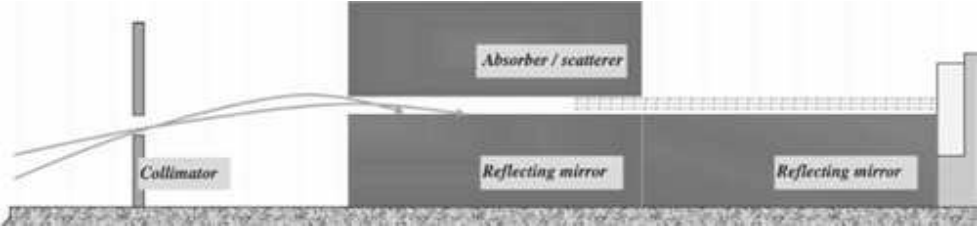}
  \caption{A schematic layout of the experiment \cite{Nev2}. A horisontal beam of neutrons
  is collimated and enters the gap between the absorber and the
  mirror. Neutrons of certain energies eventually reach the
  detector on the right, while others are absorbed.}
  \label{ExSetup}
\end{figure}

The experiment was performed with UCNs flowing between a
reflecting surface below, and an absorber above. The absorber and
the ``mirror'' create a slit through which the neutrons pass,
eventually reaching a detector at the other end of the
experimental setup (Fig. \ref{ExSetup}). UCNs are essential to the
experiment since they offer many advantages. For instance, they
have an energy of about $10^{-7} \ \mathrm{eV}$, corresponding to
a wavelength of $\sim $500 \AA \ or a velocity of $\sim $10 m/s,
allowing them to undergo total reflection at all angles against a
number of materials. The low energy also allows for high
resolution, and since neutrons have a lifetime of the order of 900
s, it is possible to store them for periods of 100 s or more. This
makes UCNs available and suitable for research and experiments in
fundamental physics. The leading research on UCNs is conducted at
LANSCE, the Los Alamos Neutron Science Center, and at ILL, the
Laue Langevin Institute, the latter holds the current
``worldrecord'' density of 41 UCNs/cm$^3$.

\begin{figure}[!htbp]
  \centering
  \includegraphics[width=8cm]{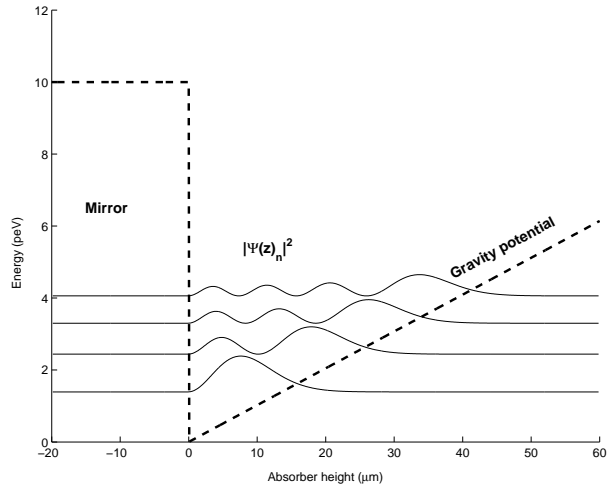}
  \caption{The energy versus the absorber height, with
  the four lowest energy eigenvalues indicated. The probability functions are plotted
  schematically. Our calculated eigenvalues are in
  accordance with the theoretical values in \cite{Nev1}.}
  \label{Gravpot}
\end{figure}

\begin{figure}[!htbp]
  \centering
  \includegraphics[width=10cm]{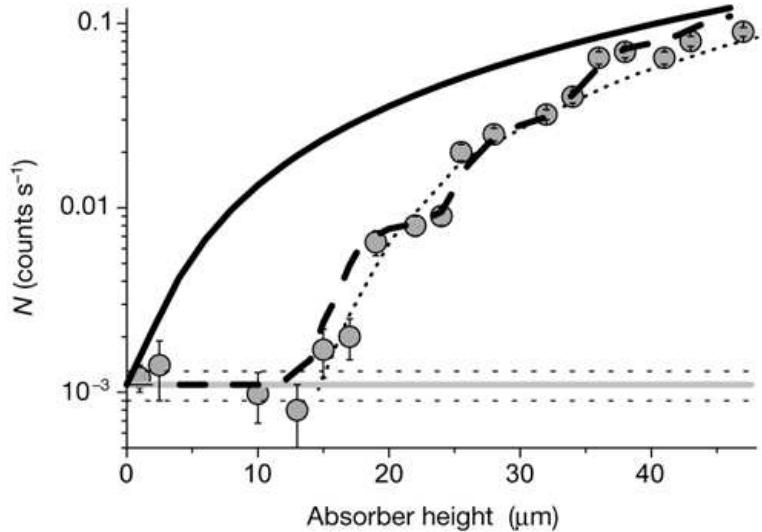}
  \caption{Neutron count versus the absorber height in the experiment
  of \cite{Nev1}. The dots with error bars show the experimental data.
  The solid line is the expectation from a "classical" analysis, while
  the dotted curve corresponds to a quantummechanical
  treatment with only the first state taken into account.
  For the dashed curve several states and level populations are considered.}
  \label{NeutronCount}
\end{figure}

Nesvizhevsky et al. argue that when the neutrons are trapped in a
potential formed by the mirror (an impenetrable "floor") and the
Earth's gravity there will be a discrete set of possible energy
levels (Fig. \ref{Gravpot}). The four lowest energy eigenvalues
are $E_1 = 1.41 \ \mathrm{peV}$, $E_2 = 2.46 \ \mathrm{peV}$, $E_3
= 3.32 \ \mathrm{peV}$ and $E_4 = 4.08 \ \mathrm{peV}$. For a
theoretical treatment of this potential, see \cite{PracQ}.
The lowest eigenvalue $E_1$ corresponds to a classical height,
$E_1=mgz$, of about 15 $\mu$m. This leads the group to predict
that when the slitopening is less than this height no neutron
transmission will occur. They argue that if the quantummechanical
wavefunction has a spatial extension larger than the opening, it
will not fit, and the neutrons have no chance of reaching the
detector. In the experiment the group observed a stepwise increase
in the number of detected neutrons as they increased the slit
height. In particular they observed, as predicted, that when the
slitopening was less than 15 $\mu$m no neutrons reached the
detector, but that there occurred a sudden increase after 15
$\mu$m and another at about 20 $\mu$m. Their results are shown in
Fig. \ref{NeutronCount}.
The curves are from the theoretical analysis. The solid line is
the fully classical treatment, where the neutron throughput does
not have a lower threshold and is described by
\begin{equation}\label{classical}
  N_{Classical}(z) \sim z^{1.5},
\end{equation}
where $z$ is the absorber height. Thus the neutron count increases
as $z^{1.5}$, since as the height is increased a larger spread of
velocities of the neutrons is allowed. The dotted line is,
according to \cite{Nev1}, the neutron count when only the lowest
eigenstate is considered. This is merely a translation of the
classical curve with the lowest threshold taken into account,
\begin{equation}\label{dotted}
  N_{\psi_1}\sim (z-z_0)^{1.5},
\end{equation}
where $z_0$ is the absorber height at which the slit becomes
transparent to neutrons. This assumes that the neutrons behave
"classically" once allowed to penetrate the gap. The dashed line
is the curve when level populations of all four lowest states were
considered.

\subsection{Reasons for further investigations}
A first thing to point out is that, the energy eigenvalues have
never been measured, i.e., all these results are entirely
theoretical. So, the only data available to us are the neutron
counts, $N$, at the detector as a function of the absorber height
$z$.

The potential used in the analysis is $V(z)=mgz$, which is the
classical gravitational potential of the Earth.

The absorber is thought to be perfect, removing the wavefunctions
(i.e., the neutrons) completely when they extend into it. From a
classical point of view, neutrons with different energies bounce
off the mirror. The different energies will result in different
heights. This is the classical explanation to why the higher
states are missing. If one then looks at the true
quantummechanical picture, the neutrons are now described by
standing waves in the gravitational potential, and there are no
"bouncing" particles. Now the absorber must also be described by a
potential and the classical argument will not be sufficient. The
absorber could hence give a geometrical explanation for the
quantum states.

The experimental statistics for larger slit widths is insufficient
\cite{Nev1}, so we need consider, say, the two first steps, making
the task of an alternative explanation a bit simpler.

\section{Alternative explanations}
In the first two subsections we try to recreate the energy
eigenvalues obtained for a particle in the Earth's gravitational
field using only geometrical arguments. However, a more careful
study of \cite{Nev1}, as well as contact with the experimental
group, makes us believe that no energy eigenvalues have actually
been measured. Instead of recreating the energy eigenvalues we
tried to explain the jumps in the number of detected neutrons.
This is presented in subsection \ref{sec: NeutronC}.

\subsection{Particle in a box}
We start with a potential consisting of two infinite walls, this
should be a fairly good approximation. The reflecting surface, the
mirror, can be seen as an infinite wall but the absorber needs
more consideration.

The problem is easy to solve analytically \cite{PracQ}
\cite{GroundQM}, and the energies are given by
\begin{equation} \label{eq: Energy eigenvalues partic in box}
  E_n = \frac{\hbar^2\pi^2n^2}{2ma^2},
\end{equation}
where $a$ is the box width. Thus, the first energy eigenvalue of a
neutron trapped in a box with infinite walls and a width of 15
$\mu$m is $E_1 = 0.9 \ \mathrm{peV}$. The first energy eigenvalue
of a neutron in the Earth's gravitational field, $E_{1grav} = 1.41
\ \mathrm{peV}$ \cite{Nev1}, is in the same range. Therefore the
assumption that we could approximate the experimental setup with a
infinite well is not quite correct, but it gives us a hint that it
might be possible to find a potential that could reproduce all the
energy eigenvalues of a neutron in a "gravitational field" with
only geometrical arguments. In the next section we try to modify
the infinite well with, a better approximation of the absorber.

\subsection{Energy eigenvalues}
The best results are obtained by using a so-called Wood-Saxon
potential, to approximate the absorber. However, the only way to
get the gravity potential results was to allow the Wood-Saxon
potential to be very close to the gravity potential itself. This
of course, was not the result we were hoping for. Consequently we
were forced to extend our discussion further. As mentioned before
we received more information about the experiment and therefore
instead of reproducing the energy eigenvalues, we only needed to
obtain the correct neutron counts and this will be described in
the next section.

\subsection{Neutron counts}\label{sec: NeutronC}
In order to reproduce the count rate we had to consider the length
(10 cm) of the experimental setup. We further assumed that a
neutron traversing the experimental setup is absorbed as
\begin{equation} \label{eq: Neutron count law}
  N(x,z) = N_{max}(z)e^{-k(z)x},
\end{equation}
where $N_{max}(z)$ is the neutron density at the entrance $x = 0$,
$x$ is the distance along the mirror, and $k(z)$ is an absorber
parameter related to how much the neutron wavefunction is inside
the absorber. The crucial condition that must be satisfied is of
course, that when the distance $z$ is less than 15 $\mu$m no
neutrons should be able to reach the detector, i.e., $N(x,z)$ must
be close to zero.

In a distance of $\Delta x$, an amount of $1-A(z,\Delta x)$
neutrons will be absorbed, yielding
\begin{equation}
  N(x_2,z) = (1 - A(z,\Delta x))N(x_1,z).
\end{equation}
Hence we get
\begin{equation}
  A(z,\Delta x) = 1 - e^{-k(z)\Delta x},
\end{equation}
where $A(z,\Delta x)$ is the area of the probability functions
inside the absorber. The function $k(z)$ can be obtained from the
experimental data with the help of Equation (\ref{eq: Neutron
count law}), giving
\begin{equation}
  k(z) = -\frac{1}{L}\ln \left[\frac{N_{out}(x,z)}{N_{max}(z)}\right],
\end{equation}
where $L$ is the length of the cavity. If we consider only the
lowest state $\psi_1(z)$, we must recreate $N_{out}(z,x)$ as the
dotted curve in Fig. \ref{NeutronCount}.\footnote{If the
transverse neutron temperature is 20 nK as stated in
\cite{Schwarz}, corresponding to $\sim 1 \ \mathrm{peV}$, even the
simple infinite box potential can explain the first step. The
smallest separation ($a \simeq 15 \ \mu$m) corresponds to the high
energy "tail" of the transverse neutron energy.} See Appendix
\ref{Appendix: Simulation} for a simulation considering only the
first state. On the other hand, when using the four lowest states,
the outcome must be the dashed line in Fig. \ref{NeutronCount}.
This case yields the probability function
\begin{equation}
  |\psi|^2 = C_1|\psi_1|^2 + C_2|\psi_2|^2 + C_3|\psi_3|^2 + C_4|\psi_4|^4,
\end{equation}
with the normalisation condition
\begin{equation}
  C_1 + C_2 + C_3 + C_4 = 1.
\end{equation}
Since we do not know the level population parameters $C_i$, we
have the freedom to choose them to make our model agree with the
experimental data.

\subsection{Combination}
There is also the possibility that the gravity potential indeed
effects the quantum states of the neutrons in this experiment.
However we would like to, unlike the group conducting the
experiment, consider the effects from the absorber as well. This
only changes the appearance of the potential in Fig.
\ref{Gravpot}, but not our principle model.

\section{Conclusions and further investigations}
The main problem in our attempts to explain the results with a
geometrical model is that $N_{max}(z)$ is unknown, depending on
the spread of energies in the neutron beam. Other difficulties are
finding realistic potentials describing the absorber and that, due
to a lack of data, our treatment is static. The time independence
will exclude important phenomena such as tunnelling effects.

Hence our results are inconclusive. We have not yet determined a
potential satisfying our requirements and we await the next report
from the experimental group, hoping it will contain the
information we need.

We propose the following improvements of the experiment:
\begin{itemize}
  \item Rotating the experimental setup by $90^\circ$ keeping
  everything else, especially the transverse neutron energies,
  constant. If the same result occurs it would indicate that the
  result is due only to the geometry of the experimental setup as
  no gravitational quantum states can form in this case.
  \item Increasing the length of the cavity. If the output would decrease
  it might confirm our theory of an absorption per unit length.
  \item A measurement of where the neutrons strike the detector. The probability
  distribution should reflect $|\psi|^2$ (since there is a standing neutron wave, the neutron is not falling
  in a classical sense).
\end{itemize}

\begin{appendix}
\section{Simulation of a potential} \label{Appendix: Simulation}
\begin{figure}[!htbp]
  \centering
  \includegraphics[width=10cm]{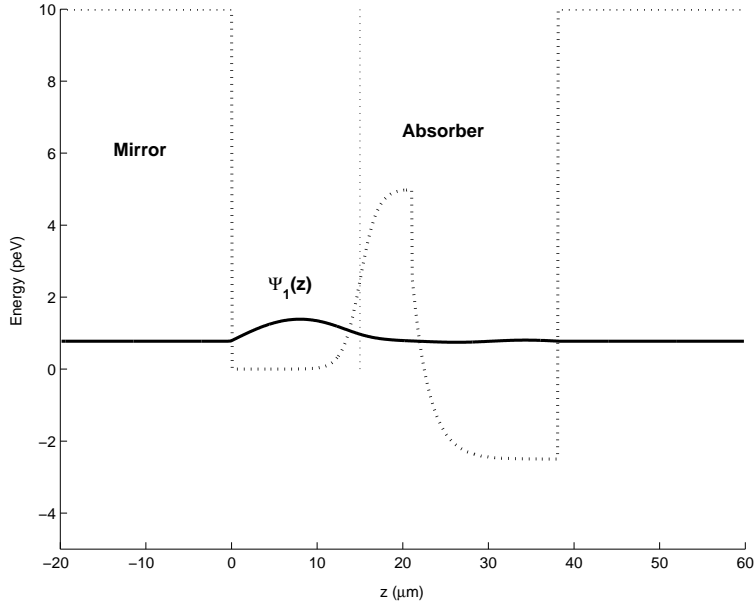}
  \caption{A simulation of a potential describing the absorber with
  the output of the first eigenfunction.}
  \label{Simulation15}
\end{figure}

We start by simulating a potential describing the absorber,
considering only the lowest state. With our algorithm we obtain
the first eigenfunction (Fig. \ref{Simulation15}) for $z = 15 \
\mu$m.
Then we calculate the normalised probability function and its area
inside the absorber,
\begin{equation*}
  A(15 \ \mu \textrm{m}, \Delta x) \simeq 0.0173 \ \textit{\textrm{area
  units}}.
\end{equation*}
By assuming $L = 10$ cm and $N_{max}(z) = 0.3$ the function $k(15
\ \mu \textrm{m})$ can be determined as
\begin{equation*}
  k(15 \ \mu \textrm{m}) \simeq 0.54991,
\end{equation*}
where we have obtained $N_{out}(10 \ \textrm{cm},15 \ \mu
\textrm{m})$ from the experimental data. Finally we can determine
$\Delta x$ as
\begin{equation*}
  \Delta x \simeq 0.0320259 \ \textrm{cm}.
\end{equation*}
Now we compare the theoretical $A(z,\Delta x)$, obtained from the
experimental data, with the simulated areas which we get with our
algorithm, see Table \ref{tab: Simu}.
\begin{table}[!htbp]
  \begin{center}
    \begin{tabular}{|c|c|c|}
      \hline
      Slit width ($\mu \textrm{m}$) & Simulated area (a.u.)& Theoretical area (a.u.)\\
      \hline
      $20$ & $0.0252$ & $0.01245$\\
      \hline
      $30$ & $0.0031$ & $0.08333$\\
      \hline
    \end{tabular}
  \end{center}
\caption{A comparison of the areas obtained from the theory versus
the simulated ones.} \label{tab: Simu}
\end{table}
The simulation and comparison of the areas show the difficulties
we had with our model. With the assumption that $N_{max}(z) =
0.3$, our area inside the absorber decreases faster than the
theoretical one when $z$ is increased. We must find a suitable
potential describing the absorber so that the two areas are the
same for all $z$. Also, the cutoff below $z = 15\ \mathtt{\mu}$m
must be satisfied.
\end{appendix}


\providecommand{\bysame}{\leavevmode\hbox
to3em{\hrulefill}\thinspace}
\providecommand{\MR}{\relax\ifhmode\unskip\space\fi MR }
\providecommand{\MRhref}[2]{%
  \href{http://www.ams.org/mathscinet-getitem?mr=#1}{#2}
} \providecommand{\href}[2]{#2}

\end{document}